\documentclass[a4paper, pra, twocolumn,superscriptaddress,showpacs]{revtex4}

\usepackage{amssymb}
\usepackage{amsmath}
\usepackage{epsfig}
\usepackage{color}
\usepackage{graphics, graphicx}
\usepackage{bbold}
\usepackage{psfrag}
\usepackage{mathcomp}
\usepackage{subfigure}
\usepackage{verbatim}
\usepackage{color}
\usepackage[colorlinks,citecolor=blue]{hyperref}

\begin{document}

\date{\today}
\title{Vortex-core structure in a mixture of Bose and Fermi superfluids}
\author{Jian-Song Pan}
\affiliation{Key Laboratory of Quantum Information, University of Science and Technology of China, CAS, Hefei, Anhui, 230026, China}
\affiliation{Synergetic Innovation Center of Quantum Information and Quantum Physics, University of Science and Technology of China, Hefei, Anhui 230026, China}
\author{Wei Zhang}
\email{wzhangl@ruc.edu.cn}
\affiliation{Department of Physics, Renmin University of China, Beijing 100872, China}
\affiliation{Beijing Key Laboratory of Opto-electronic Functional Materials and Micro-nano Devices,
Renmin University of China, Beijing 100872, China}
\author{Wei Yi}
\email{wyiz@ustc.edu.cn}
\affiliation{Key Laboratory of Quantum Information, University of Science and Technology of China, CAS, Hefei, Anhui, 230026, China}
\affiliation{Synergetic Innovation Center of Quantum Information and Quantum Physics, University of Science and Technology of China, Hefei, Anhui 230026, China}
\author{Guang-Can Guo}
\affiliation{Key Laboratory of Quantum Information, University of Science and Technology of China, CAS, Hefei, Anhui, 230026, China}
\affiliation{Synergetic Innovation Center of Quantum Information and Quantum Physics, University of Science and Technology of China, Hefei, Anhui 230026, China}

\begin{abstract}
We study a single quantized vortex in the fermionic component of a mixture of Fermi superfluid and Bose-Einstein condensate. As the density ratio between the boson and the fermion components is tuned, we identify a transition in the vortex-core structure, across which fermions in the vortex core become completely depleted even in the weak-coupling Bardeen-Cooper-Schrieffer regime. This is accompanied by changes in key properties of the vortex state, as well as by the localization of the Bose-Einstein condensate in the vortex core. The transition in the vortex-core structure can be experimentally probed in Bose-Fermi superfluid mixtures by detecting the size and visibility of the vortices.
\end{abstract}

\maketitle

\section{Introduction}
Superfluidity is one of the most remarkable macroscopic quantum phenomena. In liquid helium, and more recently in ultracold atomic gases, both Bose and Fermi superfluidity has been investigated in great detail~\cite{becsf1,becsf2,becsf3,becsf4,becsf5,fermisf1,fermisf2,fermisf3,BECbook}. A long-sought goal in liquid helium is the simultaneous superfluidity of the bosonic $^4$He and the fermionic $^3$He, which turned out to be difficult due to the strong interactions between the two species~\cite{Tuoriniemi2002,Rysti2012}. The recent experimental realization of superfluid mixtures in cold gases of Bose and Fermi atoms represents a significant step forward~\cite{Barbut2014,Yao2016,Ikemachi2016,Roy2016}. With highly tunable parameters~\cite{Chin2010}, cold atomic gases offer a flexible platform on which various many-body properties of the two-species superfluid can be investigated from new perspectives, and in a controlled fashion.

An important signature of superfluidity is the existence of quantized vortices, which have been extensively investigated in a pure Bose or Fermi superfluid of cold atoms~\cite{becsf4,fermisf1,fermisf2,BECbook}. In a very recent experiment, vortices have also been generated in a Bose-Fermi mixture of $^6$Li and $^{41}$K atoms~\cite{Yao2016}. Besides confirming the two-species superfluidity of the system, the experiment raises the challenging question regarding the vortex structure in the presence of a two-species superfluid. In particular, as the microscopic vortex-core structure offers vital information of the many-body environment~\cite{Machida2005, Chien2005, Sensarma2006} and has an important impact on the macroscopic structure of the vortex lattice~\cite{Jiang2016}, a better understanding of a single quantized vortex in the Bose-Fermi mixture is in order.

In this work, we study a singly quantized vortex in the fermionic component of the mixture of a Bose-Einstein condensate (BEC) and a Fermi superfluid. By self-consistently solving the Bogoliubov-de Gennes (BdG) and the Gross-Pitaevskii (GP) equations, we show that, as the density ratio between the boson and the fermion components increases, a transition occurs in the vortex-core structure, beyond which fermions in the vortex core become completely depleted even in the Bardeen-Cooper-Schrieffer (BCS) regime. The local depletion of fermions is accompanied by the localization of the BEC in the vortex core. The transition originates from the repulsive Bose-Fermi interactions, which induce effectively potentials of different signs for the Bose and the Fermi components. As the ground state of the many-body system changes across the transition, relevant properties such as the superfluid order parameters, the quasi-particle spectrum, and the superfluid current density of the fermions are all affected. While the transition of the vortex core persists as the Fermi superfluid is far away from the BCS regime, multiple branches of vortex bound states emerge from the bulk spectrum due to the strong extension of the vortex core. Whereas vortices in two-species BECs have been studied previously ~\cite{Matthews1999, Anderson2000, Ho1996,Skryabin2000,Ripoll2000,Garcia2000,Jezek2001,Chui2001,Catelani2010,Mason2013,Ivashin2015}, the impact of the vortex-core transition on quantities like Fermi superfluid order parameter and quasiparticle spectra is an interesting new element in the Bose-Fermi mixture, and has not been reported before.
The depletion of vortex-core fermions and the localization of BEC should be detectable under current experimental conditions by probing the size and visibility of the vortices.

The remainder of the paper is organized as follows. In Sec.~\ref{sec:model}, we present the model Hamiltonian and the equations of motion of the quasiparticles. The transition in the vortex-core structure is then discussed in Sec.~\ref{sec:transition}. The underlying mechanism of such a transition, as well as the associated changes in density distributions and quasiparticle excitations are analyzed in Sec.~\ref{sec:density}. We study the effects of interaction in Sec.~\ref{sec:BCS-BEC}, and summarize in Sec.~\ref{sec:summary}.

\section{Formalism}
\label{sec:model}
The Hamiltonian of an interacting Bose-Fermi mixture can be written as~\cite{Zheng2014,Tylutki2016}
\begin{equation}
\begin{split}
H=&\int d^{3}\boldsymbol{r}\biggl[\sum_{\sigma}\hat{\psi}_{\sigma}^{\dagger}\left(-\frac{\hbar^{2}\nabla^{2}}{2m_{F}}+g_{BF}\hat{\varphi}^{\dagger}\hat{\varphi}-\mu_{F}\right)\hat{\psi}_{\sigma}\\
&+g_{F}\hat{\psi}_{\uparrow}^{\dagger}\hat{\psi}_{\downarrow}^{\dagger}\hat{\psi}_{\downarrow}\hat{\psi}_{\uparrow}+\hat{\varphi}^{\dagger}\left(-\frac{\hbar^{2}\nabla^{2}}{2m_{B}}-\mu_{B}\right)\hat{\varphi}\\
&+\frac{g_{B}}{2}\hat{\varphi}^{\dagger}\hat{\varphi}^{\dagger}\hat{\varphi}\hat{\varphi}\biggr],
\end{split}
\label{eqn:hamiltonian}
\end{equation}
where $\hat{\psi}_{\sigma}\left(\boldsymbol{r}\right)$ is the field operator for fermions with pseudospin $\sigma=\uparrow \downarrow$ and $\hat{\varphi}\left(\boldsymbol{r}\right)$ is the field operator for bosons. $\mu_F$ ($\mu_B$) is the chemical potential of the Fermi (Bose) component. The interaction parameters for bosons are given by $g_B=4\pi\hbar^{2}a_B/m_B$ and $g_{BF}=4\pi\hbar^{2}a_{BF}/m_{BF}$, where $a_B$ and $a_{BF}$ are respectively the Bose-Bose and the Bose-Fermi scattering lengths. Here, the reduced mass $m_{BF}$ is related to the mass of bosons $m_B$ and that of fermions $m_F$ as $m_{BF}=(m_B+m_F)/(2m_B m_F)$. We consider the case where the mixture is prepared near a wide $s$-wave Feshbach resonance between the two fermion species, so that $g_F$ is related to the Fermi-Fermi scattering length $a_F$ through $1/(k_F a_F)=8\pi\epsilon_F/(g_Fk^3_F)+(2/\pi)\sqrt{E_c/\epsilon_F}$, with the Fermi energy $\epsilon_F=\hbar^2 k_F^2/2m_F$ and the Fermi wave vector $k_F$. $E_c$ is the cutoff energy introduced in the renormalization process and does not affect the results~\cite{Sensarma2006}.

Following the standard mean-field formalism, we define, respectively, the order parameter for the Fermi pairing superfluid $\Delta\left(\boldsymbol{r}\right)=g_{F}\langle\psi_{\downarrow}\psi_{\uparrow}\rangle$, and for the BEC $\varphi\left(\boldsymbol{r}\right)=\langle\hat{\varphi}\left(\boldsymbol{r}\right)\rangle$. From Eq.~(\ref{eqn:hamiltonian}), we derive the generalized BdG equation
\begin{align}
\left[\begin{array}{cc}
\hat{M} & \Delta\left(\boldsymbol{r}\right)\\
\Delta^{\ast}\left(\boldsymbol{r}\right) & -\hat{M}
\end{array}\right]\left[\begin{array}{c}
u_{n}\\
\upsilon_{n}
\end{array}\right]=E_{n}\left[\begin{array}{c}
u_{n}\\
\upsilon_{n}
\end{array}\right],
\label{eqn:BdG}
\end{align}
which is coupled to the GP equation
\begin{align}
\left[-\frac{\hbar^{2}\nabla^{2}}{2m_{B}}+g_{BF}n_{F}(\boldsymbol{r})+g_{B}n_{B}(\boldsymbol{r})\right]\varphi(\boldsymbol{r})=\mu_{B}\varphi(\boldsymbol{r}).\label{eqn:GP}
\end{align}
Here, $\hat{M}=-\frac{\hbar^{2}\nabla^{2}}{2m_{F}}+g_{BF}n_{B}\left(\boldsymbol{r}\right)-\mu_{F}$, $\left( u_{n} , \upsilon_{n}\right)^{T}$ is the fermion quasi-particle wave function with eigenvalue $E_n$, and $n_{F}\left(\boldsymbol{r}\right)=\sum_{\sigma}\langle\psi_{\sigma}^{\dagger}\psi_{\sigma}\rangle$ and $n_{B}\left(\boldsymbol{r}\right)=\left|\varphi\left(\boldsymbol{r}\right)\right|^{2}$ are respectively the fermion and the boson densities.

We study an isolated vortex state in a cylindrically symmetric trap with an open boundary condition at $\rho=R$, and periodic boundary conditions at $z=\pm L_z/2$, where $\boldsymbol{r}=(\rho, \theta, z)$ under the cylindrical coordinates. We further assume $R\gg L_z$ such that the dynamic degrees of freedom of the BEC along the $z$ direction can be neglected. To describe a single vortex, we take $\Delta(\boldsymbol{r})=\Delta(\rho)e^{-i\theta}$ and expand the quasiparticle wave functions as
\begin{equation}
\left[\begin{array}{c}
u_{n}\\
\upsilon_{n}
\end{array}\right]=\frac{1}{\sqrt{2\pi L_{z}}}\sum_{k_{z}lj}\left[\begin{array}{c}
c_{k_{z}lj}^{\left(n\right)}\phi_{jl}\left(\rho\right)\\
d_{k_{z}lj}^{\left(n\right)}\phi_{jl+1}\left(\rho\right)e^{i\theta}
\end{array}\right]e^{i(l\theta+k_{z}z)},
\label{eqn:BessExpan}
\end{equation}
where $k_z=2\nu\pi/L_z$ with $\nu\in\mathbb{Z}$. The Fourier-Bessel series $\phi_{jl}\left(\rho\right)=\sqrt{2}J_{l}\left(\alpha_{jl}\rho/R\right)/\left[RJ_{l+1}\left(\alpha_{jl}\right)\right]$ for $l\in\mathbb{Z}$ and $j\in\mathbb{Z}^{+}$. $J_{l}\left(\rho\right)$ is the Bessel function of the first kind, whose zero points are given by $\alpha_{jl}$. Similarly, we can write $\varphi\left(\boldsymbol{r}\right)$ as $\varphi\left(\boldsymbol{r}\right)=\left(2\pi L_z\right)^{-1/2}\sum_{j}f_{j}\phi_{j0}\left(\rho\right)$.

Finally, the BdG and GP equations become
\begin{equation}
\sum_{j^{'}}\left[\begin{array}{cc}
T_{l}^{jj^{'}} & \Delta_{l,l+1}^{jj^{'}}\\
\Delta_{l+1,l}^{jj^{'}} & -T_{l+1}^{jj^{'}}
\end{array}\right]\left[\begin{array}{c}
c_{k_{z}lj^{'}}^{(n)}\\
d_{k_{z}lj^{'}}^{(n)}
\end{array}\right]=E_{n}\left[\begin{array}{c}
c_{k_{z}lj}^{(n)}\\
d_{k_{z}lj}^{(n)}
\end{array}\right],
\label{eqn:BdG_FB}
\end{equation}
and
\begin{equation}
\sum_{j^{'}}\left(\frac{\alpha_{j0}^{2}}{R^{2}}\delta_{jj^{'}}+g_{BF}n_{F,0}^{jj^{'}}+g_{B}n_{B,0}^{jj^{'}}\right)f_{j^{'}}=\mu_{B}f_{j},
\label{eqn:GP_FB}
\end{equation}
where $T_{l}^{jj^{'}}=(\frac{\hbar^{2}}{2m_{F}}\frac{\alpha_{jl}^{2}}{R^{2}}+\frac{\hbar^{2}k_{z}^{2}}{2m_{F}}-\mu_{F})\delta_{jj^{'}}+g_{BF}n_{B,l}^{jj^{'}}$ with $n_{B,F,l}^{jj^{'}}=\int_{0}^{R}d\rho\rho n_{B,F}(\rho)\phi_{jl}\left(\rho\right)\phi_{j'l}\left(\rho\right)$, and
\begin{eqnarray}
\label{eqn:Delta}
\Delta_{ll^{'}}^{j^{'}j}=\int_{0}^{R}d\rho\rho\Delta\left(\rho\right)\phi_{jl}\left(\rho\right)\phi_{j'l^{'}}\left(\rho\right).
\end{eqnarray}
Equations (\ref{eqn:BdG_FB}) and (\ref{eqn:GP_FB}) are then solved self-consistently with the gap equation
\begin{equation}\label{eq:GapEQ}
\Delta\left(\rho\right)=\frac{g_{F}}{2\pi L_{z}}\sum_{lk_{z},E_{n}\geq0,jj^{'}}c_{k_{z}lj}^{\left(n\right)}d_{k_{z}lj^{'}}^{\left(n\right)}\phi_{jl}\left(\rho\right)\phi_{j^{'}l+1}\left(\rho\right),
\end{equation}
and the number equations $N_{F}=2\sum_{lk_{z},E_{n}\geq0,j}d_{k_{z}lj}^{\left(n\right)2}$ and $N_B=\sum_j f_j^2$. For our numerical calculations, we simultaneously fix the total number of particles of the fermion ($N_F$) and the boson ($N_B$) component.

\section{Transition in the vortex-core structure}
\label{sec:transition}
In the presence of the BEC, the vortex-core structure in the Fermi condensate can be significantly modified by the Bose-Fermi interactions. In current experiments with Bose-Fermi superfluid mixtures~\cite{Barbut2014, Yao2016, Ikemachi2016, Roy2016}, it is typically difficult to simultaneously tune the Fermi-Fermi and Bose-Fermi interactions to the strongly interacting regime. We therefore consider the case where the Fermi condensate is close to a wide Feshbach resonance, while the particle-number ratio of the Bose-Fermi mixture $\gamma_n=N_B/N_F$, rather than the Bose-Fermi interaction strength, is varied. We will first focus on the case in which the Fermi condensate is on the BCS side of the resonance, such that the impact of Bose-Fermi interactions on the vortex-core structure is the most significant. For simplicity, throughout this work, we consider a noninteracting BEC with $g_B = 0$, while we emphasize that all results remain qualitatively unchanged for BECs with attractive interactions or with weak repulsive interactions.

\begin{figure}[tbp]
\includegraphics[width=9cm]{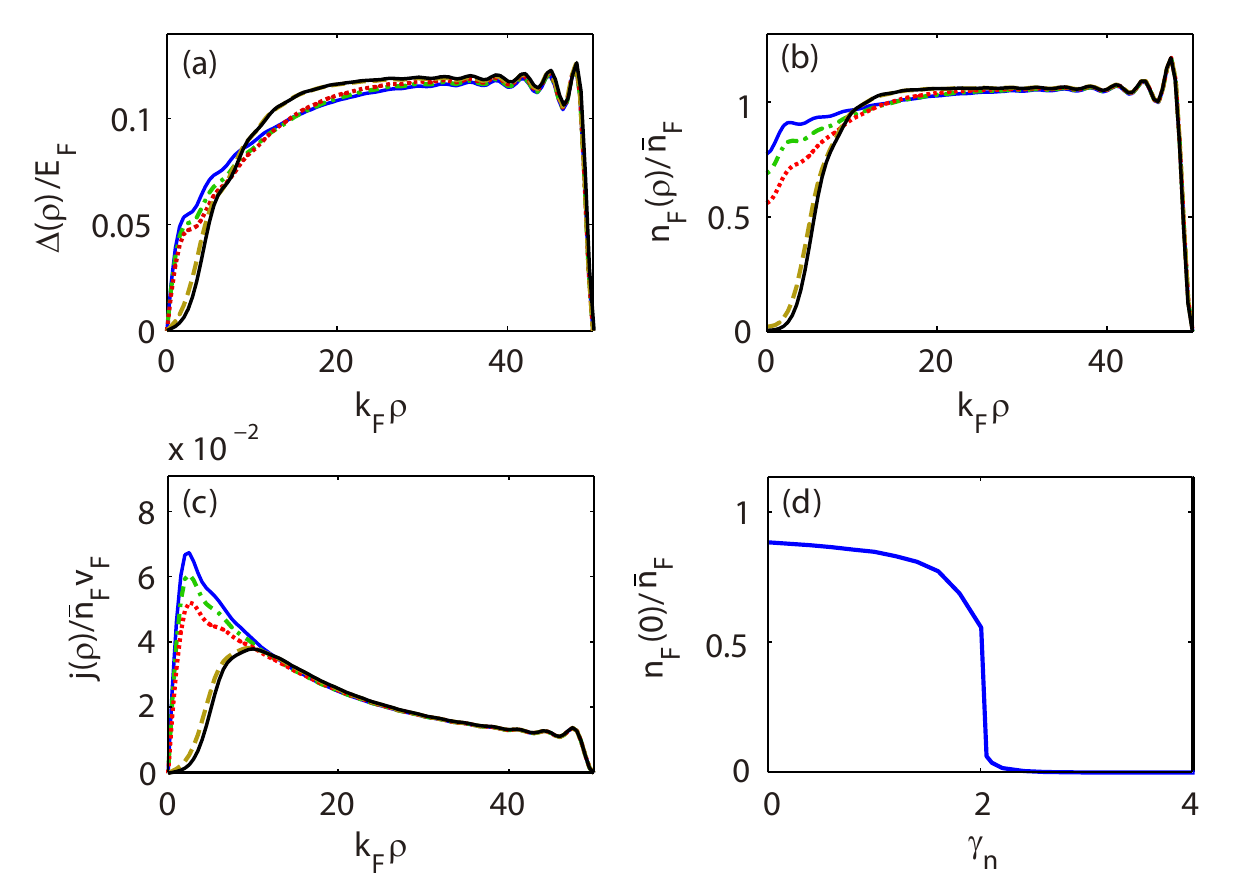}
\caption{(Color online) Order parameter (a), the fermion density distribution (b), and the superfluid current distribution (c) as functions of $\rho$ with different particle-number ratio $\gamma_n$. The blue solid, green dash-dotted, red dotted, brown dashed, and black solid curves correspond to $\gamma_n=1.6, 1.8, 2.0, 2.2, 2.4$, respectively. Here $\bar{n}_F=N_F/\pi R^2L_z$ and $v_F=\hbar k_F/m_F$. (d) The fermion density at the vortex core ($\rho=0$) as a function of $\gamma_n$. Other parameters are: $R=50/k_F$, $L_z=5/k_F$, $m_F/m_B=0.1$, $k_F=4.6\times10^6m^{-1}$, $a_F=-1/k_F$, $E_c=20E_F$, $a_B=0$, and $a_{BF}=80a_0$ with the Bohr radius $a_0$. A sharp transition occurs at around $\gamma^c_n=2.0$ for all the calculated quantities.}
\label{fig:order_parameter}
\end{figure}

In Fig.~\ref{fig:order_parameter}, we show various properties of a single vortex in the Bose-Fermi mixture as functions of $\gamma_n$. In all of these calculations, sharp transitions can be identified near $\gamma_n^{c}\approx2.0$. For $\gamma_n<\gamma_n^{c}$, the vortex-core structure has similar features as that in the absence of BEC:
the order parameter $\Delta(\rho)$ scales linearly with $\rho$ near the vortex core, the fermion
density depletes only slightly at the core, and the superfluid current distribution $j\left(\rho\right)=-2\hbar (2\pi L_{z}m_{F}\rho)^{-1}\sum_{lk_{z},E_{n}\geq0}(l+1)\left[\sum_{j}d_{k_zlj}^{(n)}\phi_{jl+1}\left(\rho\right)\right]^{2}$ has a sharp peak near $\rho_c\thicksim1/k_F$, which roughly corresponds to the size of the core~\cite{Sensarma2006,Hu2006}.
The linear dependence of the order parameter at the vortex core $\Delta(\rho) \propto \rho$ as $\gamma_n<\gamma_n^{c}$ can be semi-analytically derived by considering the asymptotic behavior of $J_l(\rho) \propto \rho^{\vert l \vert}$ as $\rho\sim0$. Close to the origin, $\Delta$ is small compared to other terms; the Bose density can therefore be seen as a constant due to its relatively large length scale of variation (see Fig. \ref{fig:size_comp_and_eff_chem}). Therefore, the BdG equation is dominated by the kinetic-energy terms and $\Delta\left(\rho\right)\propto J_{0}\left(\rho\right)J_{1}\left(\rho\right)\propto\rho$. Similarly, this initial slope of $\Delta(\rho)$ is set by the length scale $k_F^{-1}$~\cite{Sensarma2006}. Since $j=\rho_{s}v_{s}$ with $v_{s}=\left(\hbar/2m_{F}\rho\right)$ and $\rho_{s}\sim\Delta^{2}$ , $j\left(\rho\right)$ also varies linearly near the origin. When $\rho$ increases away from the vortex core, $j\left(\rho\right)\propto 1/\rho$ as $\Delta$ becomes a constant. A sharp peak in the $j(\rho)$ curve hence arises in the intermediate regime. However, as shown in Fig.~\ref{fig:order_parameter} and discussed more in the next section, when $\gamma_n$ increases beyond $\gamma_n^{c}$, these features change dramatically: the fermion density near the core center $n_F(0)$ becomes completely depleted, the order parameter deviates from a linear scaling in $\rho$ at around the core and the sharp feature in the superfluid current distribution disappears.

The occurrence of the transition at $\gamma_n^c$ is a direct result of Bose-Fermi interactions. From Eqs.~(\ref{eqn:BdG_FB}) and (\ref{eqn:GP_FB}), we see that the Bose-Fermi interactions give rise to effective potentials for the two components. While the Bose-Fermi interaction is repulsive, the effective potential near the vortex core is repulsive for fermions and attractive for bosons, which is due to the qualitatively different density distributions of $n_{B}(\rho)$ and $n_F(\rho)$. The increase of $\gamma_n$ would lead to a stronger repulsive potential $g_{BF}n_{B}(\rho)$ for the fermions, which further depletes fermions from the vortex core and increases the depth and width of the attractive potential $g_{BF}n_{F}(\rho)$ for the BEC. Such a positive feedback mechanism would eventually make the widths of the attractive and the repulsive potentials match with one another, at which point the vortex-core structure for fermions undergoes drastic changes as the BEC becomes localized at the vortex core.

\section{Density distributions and quasi-particle excitations in the vortex core}
\label{sec:density}
\begin{figure}[tbp]
\includegraphics[width=9cm]{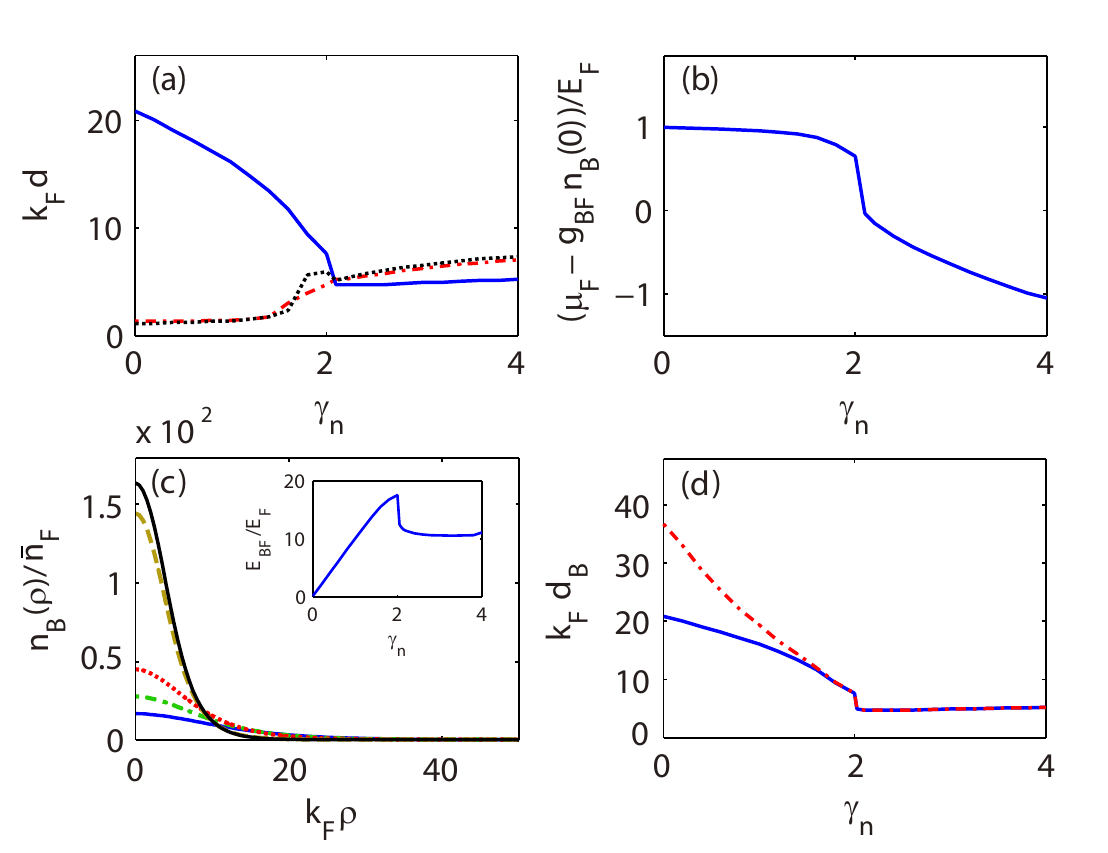}
\caption{(Color online) (a) Variation of the half-width of $n_B(\rho)$ ($d_{B}$, blue solid curve), and the half-widths of depletion in $\Delta(\rho)$ ($d_\Delta$, red dash-dotted curve) and $n_F(\rho)$ ($d_{F}$, black dotted curve). The slight distortion on the $d_F$ curve originates from the half-width position passing through a Friedel-oscillation-like peak~\cite{Machida2005}. (b) The variation of the local effective chemical potential $(\mu-g_{BF}n_B(0))$ in the vortex core. (c) The BEC density distribution with different $\gamma_n$, where the line-style convention is the same as those in Fig. \ref{fig:order_parameter}(a). The inset here shows the variation of the Bose-Fermi interaction energy $E_{BF}$ as $\gamma_n$ increases. (d) The half-widths of the BEC density distribution with $R=50/k_F$ (blue solid curve) and $R=100/k_F$ (red dash-dotted curve), respectively. Other parameters are also set to be the same as Fig. \ref{fig:order_parameter}. }
\label{fig:size_comp_and_eff_chem}
\end{figure}

The general picture above can be confirmed in Fig.~\ref{fig:size_comp_and_eff_chem}(a), where we estimate the widths of the repulsive and the attractive potentials as the half-width of $n_B(\rho)$ and the half-width of the depletion in $n_F(\rho)$, respectively. While the two widths are quite different at small $\gamma_n$, they cross each other at $\gamma_n^c$ with a length scale of $\sim 5/k_F$, and remain on the same length
scale beyond the transition. We notice that the half-width of the depletion in $\Delta(\rho)$ is also close to $5/k_F$ at $\gamma_n^c$. Although different parameters have been employed to characterize the size of the vortex core in the previous literature~\cite{Hu2006, Sensarma2006}, for convenience, here we use the half-width of fermion density depletion $d_F$ to estimate the size of the vortex core. From Fig.~2(a), it is clear that the size of the vortex core is on the order of $1/k_F$ for small $\gamma_n$, but becomes much larger beyond the transition.

\begin{figure}[tbp]
\includegraphics[width=9cm]{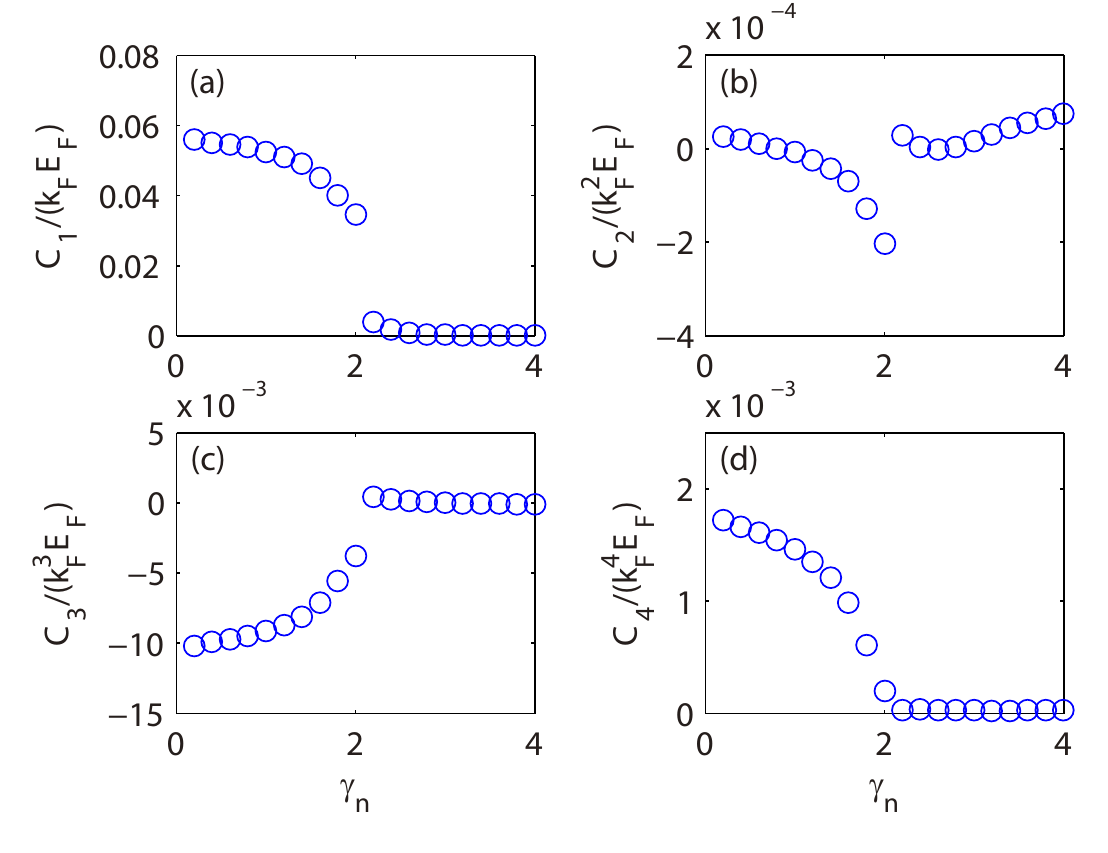}
\caption{(Color online) Coefficients $C_{j}$ of the polynomial fitting $\Delta(\rho)\approx\sum_{j=1}^{4}C_{j}\rho^{j}$ in the regime $\rho\in(0,1/k_F]$ as functions of $\gamma_n$. All parameters are set to be the same as in Fig. \ref{fig:order_parameter}. }
\label{fig:fitting}
\end{figure}

An interesting property of the transition is the complete depletion of fermions at the center of the vortex. The fermion depletion can be understood by calculating the effective chemical potential $\mu_{\rm eff}(\rho)=\mu_F-g_{BF}n_B(\rho)$ of the fermions at $\rho=0$. When $\gamma_n>\gamma_n^c$, the variation length scale of the fermion density at the vortex core becomes several times of $1/k_F$ and the local density approximation (LDA) should be locally applicable. As the local pairing order parameter vanishes at $\rho=0$, according to the number equation for fermions, the local fermion density at $\rho=0$ should also vanish when the local chemical potential $\mu_{\rm eff}$ becomes negative. We show $\mu_{\rm eff}(0)$ as a function of $\gamma_n$ in Fig.~\ref{fig:size_comp_and_eff_chem}(b), from which it is clear that the effective chemical potential undergoes a sharp transition at $\gamma_n^c$ and becomes negative as $\gamma_n>\gamma_n^c$. Numerically, we find the fermion density near the core center becomes vanishingly small when $\gamma_n>\gamma_n^c$. The complete depletion of $n_F(\rho)$ and the localization of the bosons discussed below can be seen as a phase separation~\cite{Tylutki2016}. Accordingly, $\Delta(\rho)$ and $j(\rho)$ also vanish near the core center, which leads to the deviation of the order parameter from a linear scaling in $\rho$ and the disappearance of sharp features in the superfluid current distribution. This can be confirmed through a polynomial fitting of $\Delta(\rho)$ near the core center, where the fitting coefficients present sharp features at the transition point and become vanishingly small as $\gamma_n>\gamma_n^c$, as shown in Fig.~\ref{fig:fitting}.

The depletion of the fermions at the core is accompanied by the localization of the BEC wave function for $\gamma_n>\gamma_n^c$. Such a localization can be directly identified from the density distribution of the BEC as shown in Fig.~\ref{fig:size_comp_and_eff_chem}(c), and is consistent with the variation of the half-width of its density distribution in Fig.~\ref{fig:size_comp_and_eff_chem}(a). When $\gamma_n$ is small, the fermion depletion is limited both in particle number and in spatial range. The BEC density distribution is therefore extended, with the half-width comparable to the system size $R$. When $\gamma_n$ increases, the half-width decreases continuously, suggesting a stronger attractive potential provided by the vortex state and a localizing tendency of the extended BEC. Near the critical $\gamma_n^c$, the half-width of the boson density distribution is $\sim 5/k_F$, which is comparable to the size of the vortex core. By further increasing $\gamma_n$ beyond $\gamma_n^c$, the half-width begins to increase with $\gamma_n$, which is a clear signal that the BEC is localized in the core. After the localization of the BEC and the depletion of the fermions from the core, the Bose-Fermi interaction energy mostly comes from the Bose-Fermi boundary, which gives rise to a saturation of the interaction energy beyond $\gamma_n^c$ [see the inset of Fig.~\ref{fig:size_comp_and_eff_chem}(c)]. The localization of the BEC is also reflected in the dependence of the half-width of the BEC density distribution on the system size $R$.
As illustrated in Fig.~\ref{fig:size_comp_and_eff_chem}(d), while the half-width increases significantly with $R$ for $\gamma_n<\gamma_n^c$, beyond the transition, the half-width is essentially insensitive to $R$.

\begin{figure}[tbp]
\includegraphics[width=9cm]{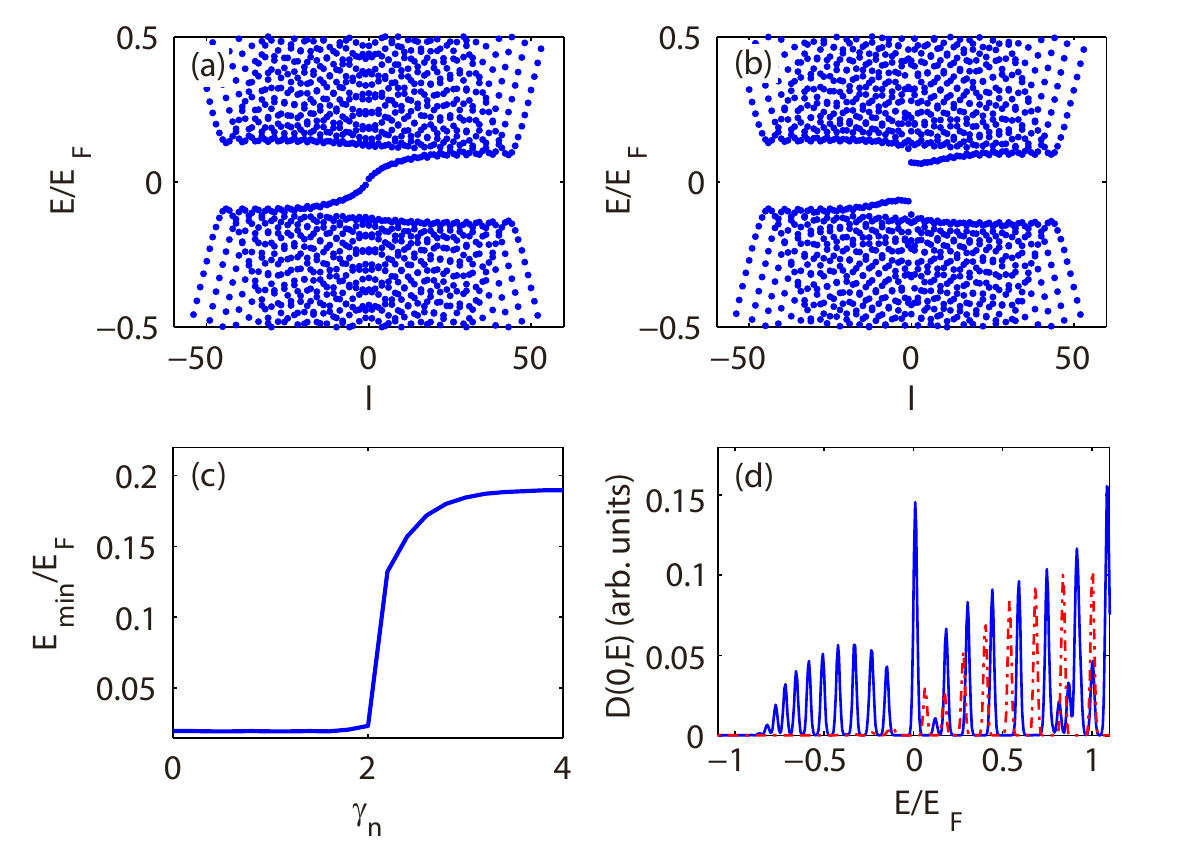}
\caption{(Color online) The Fermi quasiparticle spectra for $\gamma_n=1.8$ (a) and $2.2$ (b), and the variation of the energy of the lowest excited state (c) as functions of $\gamma_n$. (d) The local density of states $D(0,E)$ for $\gamma_n=1.8$ (blue solid curve) and $2.2$ (red dash-dotted curve). Other parameters are also the same with those in Fig. \ref{fig:order_parameter}.}
\label{fig:bound_state}
\end{figure}

The fermion depletion and the modified behavior of $\Delta(\rho)$ at the vortex core give rise to changes in the quasiparticle excitations and in the Andreev-like bound states at the vortex core. These changes are illustrated in Fig.~\ref{fig:bound_state}. Beyond the critical $\gamma_n^c$, the nearly continuous in-gap spectrum will be pushed toward the bulk spectrum [see Fig. \ref{fig:bound_state}(a)-\ref{fig:bound_state}(c)], with an appreciable increase in the lowest excitation energy for the Andreev-like bound states. The suppression of the in-gap bound states as well as the complete depletion of fermions at the core, are reflected in the local density of states (LDOS) $D(\rho,E)=2\sum_{n}|u_n(\rho)|^2\delta(E-E_n)$~\cite{Hu2006,Iskin2008} at $\rho=0$. As shown in Fig. \ref{fig:bound_state}(d), while the sharp peak associated with the lowest excited state shrinks, states with $E<0$ become unoccupied in the LDOS for $\gamma_n>\gamma_n^{c}$.

\section{BCS-BEC crossover}
\label{sec:BCS-BEC}
\begin{figure}[tbp]
\includegraphics[width=9cm]{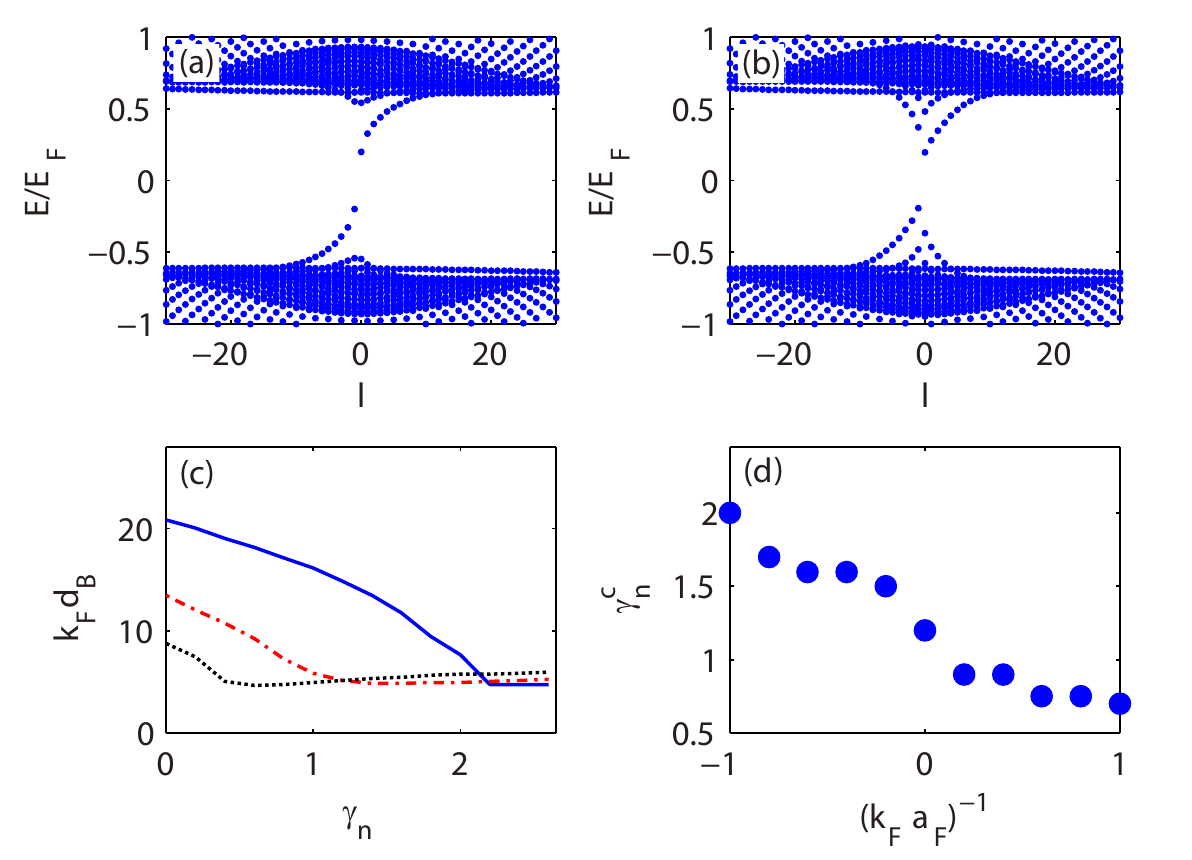}
\caption{(Color online) Quasiparticle spectra at $(k_F a_F)^{-1}=0$ with $\gamma_n=1$ (a) and $1.4$ (b). Here the critical $\gamma_n$ is $\sim 1.2$. The quasiparticle spectra in the BEC regime is qualitatively similar with those at unitarity. For comparison, we plot the half-widths of the Bose density distributions $d_{B}$ for $(k_F a_F)^{-1}=-1$ (blue solid curve), $(k_F a_F)^{-1}=0$ (red dash-dotted curve) and $(k_F a_F)^{-1}=1$ (black dotted curve) in (c). (d) The variation of the critical $\gamma_n$ as a function of $(k_F a_F)^{-1}$.}
\label{fig:BCS_BEC}
\end{figure}

In the previous discussions, we have been focusing on the vortex state in the BCS regime. As the Fermi superfluid is tuned across the BCS-BEC crossover, the transition in the vortex-core structure persists, while many features are modified. Away from the BCS regime, as the pairing order parameter becomes larger, the lowest excitation energy of the Andreev-like bound states also increases, which is similar to the case without BEC~\cite{Sensarma2006}. Additionally, unique to the Bose-Fermi superfluid mixture, multiple branches of vortex bound states emerge from the bulk spectrum away from the BCS regime [see Fig.~\ref{fig:BCS_BEC}(a) and \ref{fig:BCS_BEC}(b)], which is only possible when the core size becomes larger than the coherence length $\xi=\hbar v_F/\Delta_0$~\cite{Bardeen1969}. Here, $\Delta_0$ is the bulk value of $\Delta(\rho)$ and $\xi\backsim1/k_F$ in the resonance regime~\cite{Sensarma2006}. The condition above is facilitated by the larger fermion depletion at the vortex core in the strong-coupling regime, which enhances BEC localization and increases the core size; and by the smaller coherence length of the Fermi superfluid away from the BCS regime. We then identify the transition in the vortex core, for instance, from the half-width of the BEC density distribution [see Fig.~\ref{fig:BCS_BEC}(c)], or from features of the vortex state. As shown in Fig.~\ref{fig:BCS_BEC}(d), the critical $\gamma_n^c$ decreases as the Fermi-Fermi interaction is tuned toward the strong-coupling regime, which further confirms the enhanced BEC location toward the strong-coupling regime.

\section{Summary and Final remarks}
\label{sec:summary}

By analyzing the vortex state in the Fermi component, we have revealed an interesting transition in the vortex-core structure of a mixture of Bose and Fermi superfluids. We found that the vortex-core transition can lead to a variety of interesting features in quantities like the superfluid order parameter and the quasiparticle spectra of the fermions. In particular, we identified a partial depletion of the fermion density at the vortex core as well as the Andreev-like vortex bound states in the quasiparticle spectra. Such a transition is induced by the inter-species interaction between the two superfluid components. When the Fermi-Fermi interaction is tuned to the deep BEC regime, the vortex state discussed here can be reduced to the previously studied case of a two-species BEC~\cite{becsf3, Anderson2000, Ho1996, Skryabin2000, Ripoll2000, Garcia2000, Jezek2001, Chui2001, Catelani2010, Mason2013, Ivashin2015}, where a transition of vortex-core structure is also present due to the interspecies interaction. However, we emphasize that the interesting features discussed in this work are unique for a Bose-Fermi superfluid mixture, as a vortex in a Fermi superfluid is linked to a phase singularity of the pairing field and hence acquires much richer structures away from the deep BEC regime.

\begin{figure}[tbp]
\includegraphics[width=9cm]{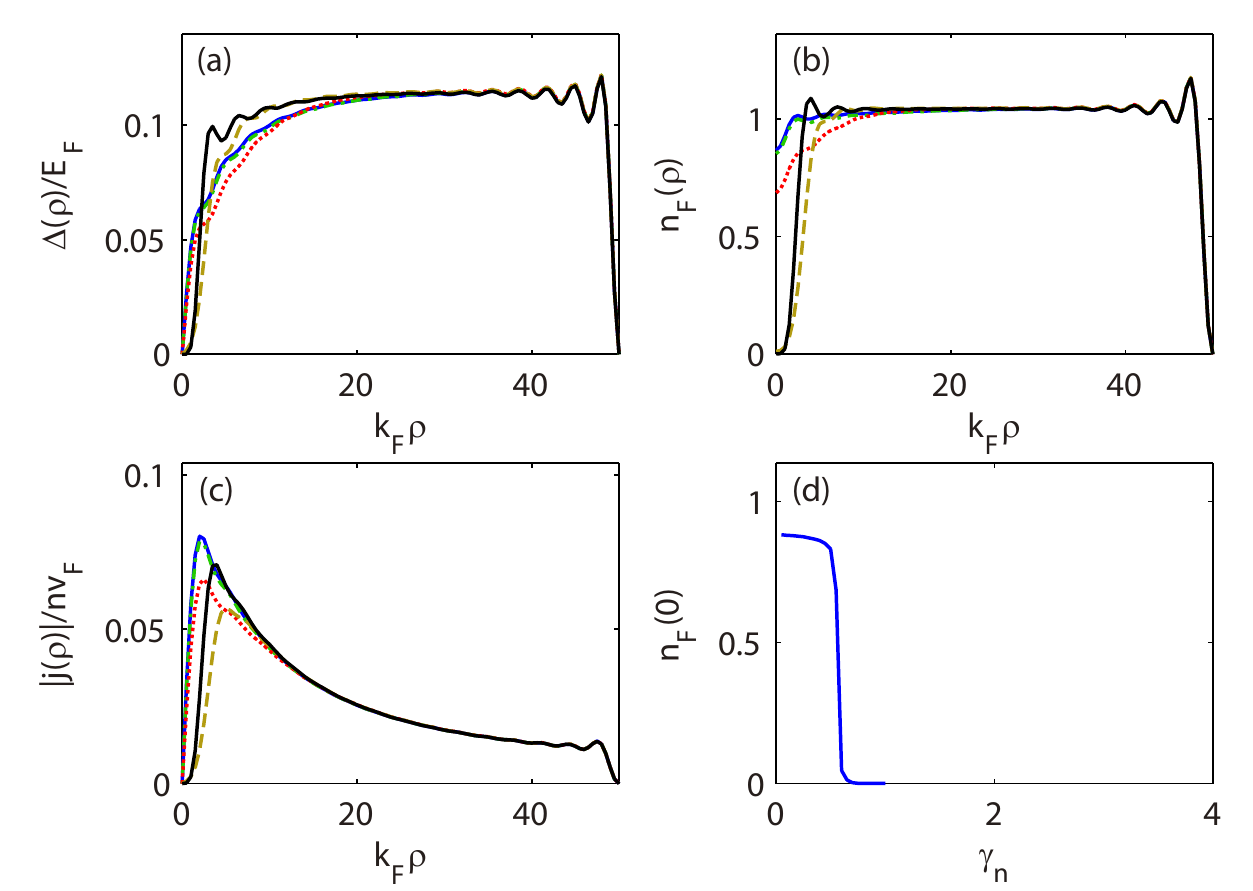}
\caption{(Color online) (a) Order parameter, (b) the fermion density distribution, and (c) the superfluid current distribution as functions of $\rho$ with different particle-number ratio $\gamma_n$. The blue solid, green dash-dotted, red dotted, brown dashed, and black solid curves correspond to $\gamma_n=0.35, 0.45, 0.55, 0.65, 0.75$, respectively.  (d) The fermion density at the vortex core ($\rho=0$) as a function of $\gamma_n$. Here, we consider a weak attractive interaction between bosons with $g_B=-10a_0$, while the other parameters are the same as in Fig. \ref{fig:order_parameter}.}
\label{fig:negg_B}
\end{figure}

Furthermore, although we have mostly considered a noninteracting BEC, we have checked that a similar transition exists for BEC with attractive interactions or with weak repulsive interactions. Strong repulsive interaction in the BEC would compete with the attractive potential of the vortex state, and thus prevent the localization of the BEC and the corresponding transition in the vortex core. We show the transition with an attractive interaction for the bosons in Fig. \ref{fig:negg_B}, where the critical $\gamma_n$ even becomes smaller. Unlike the case with $g_B=0$, the BEC is further localized due to the attractive interaction and the sharp feature on the $j(\rho)$ curve appears again as $\gamma_n$ far away from the critical point. In the case of a repulsive BEC, the parameter regime for the vortex-core transition is extremely narrow under current parameters. The transition in the vortex core should significantly change the size and visibility of the vortices, and it would be interesting to study its effects on the macroscopic vortex-lattice configurations under experimental conditions.

\section*{Acknowledgments}
We thank G. Catelani for helpful comments. This work is supported by the National Key R\&D Program (Grant No. 2016YFA0301700), the NKBRP (Grant No. 2013CB922000), the National Natural Science Foundation of China (Grants No. 60921091, No. 11274009, No. 11374283, No. 11434011, No. 11522436, and No. 11522545), and the Research Funds of Renmin University of China (Grants No. 10XNL016 and No. 16XNLQ03). W. Y. acknowledges support from the "Strategic Priority Research Program(B)" of the Chinese Academy of Sciences, Grant No. XDB01030200.

\end{document}